\documentclass[pra,twocolumn,showpacs,floatfix,amsfonts]{revtex4}

\newcommand{\eqb}{\begin{eqnarray}}
\newcommand{\eqe}{\end{eqnarray}}

\usepackage{epsfig}
\usepackage{graphicx}
\usepackage{amsmath}
\usepackage{amssymb}
\usepackage{dcolumn}
\usepackage{color}

\begin{document}

\title{Nonlinear patterns in Bose-Einstein condensates in dissipative optical lattices.}

\author{Yu. V. Bludov$^1$}
\email{bludov@fisica.uminho.pt}

\author{V. V. Konotop$^{2}$}
\email{konotop@cii.fc.ul.pt}

\affiliation{ $^1$Centro de F\'{\i}sica, Universidade do Minho, Campus de Gualtar, Braga 4710-057, Portugal
\\
$^2$Centro de F\'{\i}sica Te\'orica e Computacional,
Universidade de Lisboa, Complexo Interdisciplinar, Avenida Professor
Gama Pinto 2, Lisboa 1649-003, Portugal and Departamento de
F\'{\i}sica, Universidade de Lisboa, Campo Grande, Ed. C8, Piso 6,
Lisboa 1749-016, Portugal}

\pacs{03.75.Kk, 03.75.Lm, 67.85.Hj}

\begin{abstract}
It is shown that  the one-dimensional nonlinear Schr\"odinger equation with a dissipative periodic potential, nonlinear losses and linear pump allow for the existence of stable nonlinear Bloch states which are attractors. The model describes a Bose-Einstein condensate with inelastic two-- and three--body interactions loaded in an optical lattice with losses due to inelastic interactions of the atoms with photons.

\end{abstract}

\maketitle

\section{Introduction}

Being indispensable attribute of all natural processes the dissipation accompanied by properly adjusted compensation of losses, plays a constructive role in generation of nonlinear patterns, which nowadays are observed in many natural sciences, including physics, biology, chemistry, life sciences, etc~\cite{Akhmediev}. As attractors the respective structures play a  prominent role in practical applications among which we particularly mention light patterns in nonlinear optics, receiving a rapidly increasing deal of attention~\cite{optics}. For description of the emergence of dissipative patterns   several mathematical models have been developed, the complex Ginzburg-Landau  equation being the  most simple and widely used one.  This model  allows  for quantitative  and often even  qualitative description of a wide range of the physical phenomena~\cite{GL}.

The dissipation is also an important factor in the theory of Bose-Einstein condensates (BECs), where it appears in a natural way either through the inelastic interactions of light with atoms (see, e.g.,~\cite{Pethick}), or in a form of inelastic two-body and three-body inter-atomic interactions~\cite{FMS,nonlin-dissip_Fedichev,nonlin-dissip}. Alternatively, the dissipation can be introduced artificially,  say in a form of nonlinear dissipative lattice~\cite{Abdullaev} or by probing the condensate by means of an electronic beam~\cite{exper},  and can affect the matter wave dynamics in a variety of ways~\cite{BKPO}. It is then natural to expect that a BEC where the dissipative losses are properly compensated by a pump of atoms to the system, can give rise to highly stable atomic density distributions. Probably the simplest nontrivial and practically important realization of such situation can be a BEC loaded in a one-dimensional (1D) optical lattice (OL).

This leads us to the aim of the present work: we report theoretical study of  stable matter wave patterns emerging in an array of BECs when the linear and nonlinear dissipations are compensated by the homogeneous pump of the atoms. More specifically, we show that in such  systems emergence of {\em nonlinear dissipative Bloch waves} is possible.  Such  periodic patterns appear as attractors, and hence  initial distributions in a large range of the parameters rapidly evolve to the dissipative Bloch waves. These are essentially nonlinear states, which in a general situation do not allow for transition to the limit of zero density, and at the same time, when the dissipation is not too strong, can be naturally associated with a specific edge of the band of the underline conservative part of the lattice potential.

The organization of the present paper is as follows. In Sec.~\ref{se:model} we formulate the models of the dissipative cubic nonlinear Schr\"odinger (NLS) equation and provide the estimates for the main effects to be observable.  In the next Sec.~\ref{sec:numerics} we perform numerical study of the emergence and stability of the periodic patterns. Sec.~\ref{sec:quintic} is devoted to extension of the results to the NLS equation with quintic dissipation. In Sec.~\ref{sec:lattice} we present the respective lattice models, which can be obtained using the tight-binding approximation from the initial continuous dissipative NLS equations. The results are summarized in Conclusion.

\section{The model and analytical estimates}
\label{se:model}

\subsection{The model}

We start with the  1D Gross-Pitaevskii (GP) equation
\begin{eqnarray}
\label{eq:NSE-0}
i\Psi_t=- \Psi_{xx}+2\alpha \sin^2(x)\Psi+g |\Psi|^2\Psi+i\Gamma\Psi
\end{eqnarray}
where $\alpha=\alpha^\prime-i\alpha^{\prime\prime}$ is the properly normalized complex atomic polarizability whose imaginary part $\alpha^{\prime\prime}>0$ becomes appreciable for relatively large electric fields (see, e.g.,~\cite{Pethick}), $g=g^\prime-ig^{\prime\prime}$ is the dissipative nonlinearity (with $g^{\prime\prime}>0$ describing inelastic interatomic collisions) and   $\Gamma>0$ is the linear gain.

It is relevant to mention, that emergence of coherent localized and periodic structures from the interplay between periodicity and complex (i.e. conservative and dissipative) nonlinearity have recently being  addressed within the framework of the complex Ginzburg-Landau equation~\cite{SakMalom}  and nonlinear dissipative Bloch waves in a model of optical parametric oscillators were reported in Ref.~\cite{BKT}. Compared to these previous studies, apart from the new physical applications, the present model (\ref{eq:NSE-0}) display  very different mathematical features, which stem from the periodically varying dissipation which is controlled by the OL.

For the next consideration it is convenient to introduce the parameter
\begin{eqnarray}
\label{delta}
\Delta=\Gamma-\alpha^{\prime\prime},
\end{eqnarray}
 which will play the role of the effective gain (see below), and parameter $\phi$ defined by the relations $g^{\prime\prime}/|g|=\sin\phi$ and  $g^{\prime}/|g|=\cos\phi$, which controls the nonlinear dissipation (the parametrization introduced  implies that  $0\leq\phi\leq\pi$). Then,  renormalizing the macroscopic wavefunction
$\displaystyle{\psi =|g|^{1/2}e^{-i\alpha^\prime t}\Psi }$ we rewrite  Eq.~(\ref{eq:NSE-0}) in the form
\begin{equation}
i\psi_t=- \psi_{xx}-(\alpha^{\prime}-i\alpha^{\prime\prime})\cos(2x)\psi+e^{-i\phi}|\psi|^2\psi+i\Delta\psi.
\label{eq:NSE}
\end{equation}

\subsection{On homogeneous linear pump}
\label{subsec:lin_pump}

The model (\ref {eq:NSE}) explored in the present paper is based  on the supposition that there exists some "gain" mechanism compensating dissipative losses.  Generally speaking such a mechanism would require an infinite external reservoir of atoms supplying the condensate. Inclusion of the linear gain has been explored in the BEC applications~\cite{FMS,Lin_Diss} and is usually associated to the growth of the number of condensed atoms due to condensation from the thermal cloud~\cite{FMS}. However,  a possibility of controlled  and  {\it a priori} given gain of the number of atoms, to the best of our knowledge, has not been proposed, so far.

In this subsection we outline an idea on how  an {\em effective} homogeneous gain can be implemented  "mechanically". It  is based on the fact that, what really matters is the linear atomic density rather than the  total number of atoms. Hence, one can consider a cigar-shaped condensate whose dimensions decrease  in time, thus resulting in increase of the linear density.

To this end we consider a 1D BEC loaded into a cigar-shaped potential, which consists of a periodic OL and a parabolic trap. In order to implement an effective linear homogeneous pump we assume that (i) both the longitudinal and transverse harmonic oscillator frequency are growing functions of time: $\omega_\parallel(T)=\omega_\parallel f(T)$ and $\omega_\bot(T)=\omega_\bot f(T)$, where $\omega_\parallel$ and $\omega_\bot$ are the respective frequencies at initial time $T=0$
and $f(T)$ is  a positive definite increasing function, chosen to be  $f(T)\equiv (1-2\omega_\parallel T)^{-1}$; (ii) the complex s-wave scattering length grows in time: $a(T)=a_sf(T) $, $a_s=a_s^\prime-ia_s^{\prime\prime}$ being the initial complex scattering length (the desired time dependence can be achieved with help of Feshbach resonance), (iii) the lattice amplitude and the lattice period are functions of time: $V(T)=Vf(T) $ and $d(T)=d\, f^{-1/2}(T)$, $V=V^\prime-iV^{\prime\prime}$ and $d$ being the respective initial values (this can be achieved by changing the laser beam intensity and the angle between the beams creating the OL).  Then, in the mean-field approximation such a condensate is described by the 3D GP equation with varying coefficients:
\begin{eqnarray}
\label{eq:GP1}
i\hbar\Psi_T=-\frac{\hbar^2}{2m}\nabla^2\Psi+ 2V(T)\sin^2\left(\frac{\pi X}{d( T)}\right)\Psi
\nonumber\\
+ \frac{m}{2}\left[\omega^2_\parallel(T) X^2+\omega^2_\bot(T) R^2_\bot\right]\Psi+\frac{4\pi\hbar^2a(T)}{m}|\Psi|^2\Psi,
\end{eqnarray}
where $\Psi$ and $m$ are wavefunction (in the physical units) and the mass  of the atoms, respectively, $\textbf{R}_\bot=(Y,Z)$ is the transverse coordinate.

In the case when OL period $d$ is much bigger than the transverse harmonic oscillator length, $d \gg a_\bot=\sqrt{\hbar/(m\omega_\bot)}$, Eq.~(\ref{eq:GP1}) can be reduced to the 1D GP equation   by means of the multiple-scale expansion procedure~\cite{KS}. To this end   the 3D bosonic wavefunction has to be searched in the form
\begin{eqnarray}
\Psi(X,\textbf{R}_\bot,T)=\frac{\pi a_\bot}{2d|a_s|^{1/2}}\psi(x,t)\zeta\left(\textbf{R}_\bot,T\right)\times\nonumber\\
\exp{\left(-i\frac{ V^\prime T}{\hbar}-i\frac{\omega_\bot}{2\omega_\parallel}\log[f(T)]-i\frac{\omega_\parallel f(T)X^2}{2\omega_\bot a_\bot^2}\right)},
\label{eq:psi-3d}
\end{eqnarray}
where $\psi(x,t)$ is the  unknown dimensionless function of the new dimensionless independent variables
\begin{eqnarray}
\label{dim_less}
x=\frac{\pi}{d} \sqrt{f(T)}X \quad\mbox{ and}\quad  t = \frac{E_r}{2\hbar\omega_\parallel}\log[f(T)]
\end{eqnarray}
 (hereafter $E_r=\hbar^2\pi^2/(2md^2)$ is the "initial" recoil energy) and  $\zeta(\textbf{R}_\bot,T)=\pi^{-1/2}a_\bot^{-1}\exp\left[-R^2_\bot f(T)/(2a_\bot^2)\right]$ describes the linear transverse distribution varying in time due to change of the transverse trap. Then, introducing the  dimensionless parameters  $\alpha=\alpha^\prime-i\alpha^{\prime\prime}=(V^\prime-iV^{\prime\prime})/E_r$,   $\Delta=3\hbar\omega_\parallel/(2E_r)-\alpha^{\prime\prime}$, and  $\phi$, the latter defined by the relations $a_s^{\prime\prime}/|a_s|=\sin\phi$ and  $a_s^{\prime}/|a_s|=\cos\phi$,  and substituting the ansatz
 (\ref{eq:psi-3d}) in  Eq.~(\ref{eq:GP1}) one readily obtains the desirable equation (\ref{eq:NSE}).

As it is clear the above mechanism formally works only for $T<1/2\omega_\parallel$ and implies change of the trap dimensions, which should not affect the applicability of the meanfield model~\footnote{The consideration in this paper was restricted to 1D models, whose validity is well justified in the limit of low densities and tight transverse binding (see e.g.~\cite{KS}). The full 3D simulations accounting possible transverse effects will be published elsewhere.}. This means that the created effective pump is relatively small, and thus can properly work for small dissipative losses. Say, if in the dimensionless units $\Gamma\sim\alpha^{\prime\prime}\sim 0.001$ , for the rubidium condensate loaded   in a cigar-shaped trap with the initial dimensions $a_\bot=0.5\,\mu$m and $a_\parallel=17.32\,\mu$m (corresponding to the $\omega_\parallel\approx 2.42\,\,$Hz), and with  the imposed optical lattice with the constant  $d=1\,\mu$m, one computes that the time interval during which the constant pump can be supported is $T\approx 155\,\,$ms (or $t\approx 1040$ in dimensionless units), and corresponds to the decrease of the linear dimensions of the trap two times.

Below, in this work, however we explore more strong values of the gain (and dissipation), in order to reduce time during which  instabilities are developed and unstable initial conditions converge to the respective attractors. More specifically, following~\cite{nonlin-dissip_Fedichev} for all numerical simulation we will chose $\Gamma=0.1$.

\subsection{On the spectrum of the linear problem.}
\label{sec:lin_spec}

We are particularly interested in periodic stationary patterns. Therefore we explore the ansatz $\psi(x,t)=\varphi(x)\exp(-i\mu t)$, with a real chemical potential $\mu$ leading the nonlinear eigenvalue problem
\begin{eqnarray}
\mu\varphi=L\varphi  +i[\Delta +\alpha^{\prime\prime} \cos(2x)] \varphi+e^{-i\phi}|\varphi|^2\varphi
\label{eq:NSE_lin}
\end{eqnarray}
where we have introduced the linear operator $L=-\frac{d^2}{dx^2}-\alpha^\prime\cos(2x)$. The eigenvalues and eigenfunctions of this operator are obtained from the Mathieu equation
\begin{equation}
L\varphi_{nk} =E_n(k)\varphi_{nk}.
\label{eq:mathieu}
\end{equation}
It gives the  band-gap spectrum, characterized by the number of the band $n\ge 1$ ($n=1$ corresponding to the lowest band) and by the wave vector $k$ in the first Brillouin zone, $k\in[-1,1]$. Subsequently, the spectrum of the operator $L$ is given by $E_n(k)\in [E_1^{(-)} ,E_1^{(+)}]\cup[E_2^{(-)},E_2^{(+)}]\cup\cdots$ ($E_1^{(-)}<E_1^{(+)}<E_2^{(-)}<...$). Here $E_n^{(-)}$ and $E_n^{(+)}$ are respectively the lower and the upper edges of the $n$-th band, $E_n^{(-)}=E_n(0)$, $E_n^{(+)}=E_n(1)$, if $n$ is odd, and $E_n^{(-)}=E_n(1)$, $E_n^{(+)}=E_n(0)$, if $n$ is even.

The requirement for $\mu$ to be real readily gives the relation
\begin{equation}
\int_{0}^{\pi} \left[\alpha^{\prime\prime}\cos(2x)|\varphi |^2-\sin(\phi)|\varphi |^4\right]dx +\Delta N=0
\label{eq:N0}
\end{equation}
where $N=\int_{0}^{\pi} |\psi|^2 dx$
is the normalized number of atoms per one lattice period.  We notice, that   the   condition (\ref{eq:N0}) can also be obtained from the requirement for the number of condensed atoms to conserve, i.e. from  $\partial N/\partial t\equiv 0$.

Our analysis will be restricted to the case of weak dissipation, which we express in terms of the small parameter $\varepsilon=\sqrt{|\alpha^{\prime\prime}/\alpha^{\prime}|}\ll 1$ and respectively we require $\Delta=\varepsilon^2 \delta$ with $|\delta|\lesssim 1$. It is then natural to recall that in the absence of the dissipation Eq. (\ref{eq:NSE}) possesses branches of periodic solutions which in the linear limit, i.e. when $N\to 0$, are reduced to the conventional Bloch states
$\varphi_{nk}$. Moreover, when $N\ll 1$ the modulational instability of the respective periodic solutions is described in terms of the multiple-scale approximation~\cite{KS}. Therefore we start the description of the dissipative problem at hand with this limit.

As the first step we consider the case when in the respective  linear problem
\begin{eqnarray}
L\varphi + i \varepsilon^2 V(x)\varphi =E \varphi,
\end{eqnarray}
the dissipation $ V(x)\equiv\delta + \alpha^{\prime} \cos(2x)$  is treated as a perturbation. It is straightforward application of the perturbation theory to obtain corrections for a chosen eigenvalue $E_n(k)$ 
due to the dissipative term:
\begin{eqnarray}
E=E_n(k)+\varepsilon^2 E_{nk}^{(1)} +\varepsilon^4 E_{nk}^{(2)} +\cdots
\end{eqnarray}
 with
\begin{eqnarray}
E_{nk}^{(1)}=i\int^{\pi}_0{ V(x)|\varphi_{nk}(x)|^2 dx}
\end{eqnarray}
and
\begin{eqnarray} \displaystyle E_{nk}^{(2)}=-\sum_{n^\prime\neq n}\frac{\left|\int^{\pi}_0{ V(x)\varphi_{nk}(x) \overline{\varphi}_{n^\prime k}(x) dx}\right|^2}{E_n(k)-E_{n^\prime}(k)}
\end{eqnarray}
(hereafter an overbar stands for the complex conjugation).
 The obtained result has several important consequences. Indeed, for the eigenvalue to be real, one has to require $E_{nk}^{(1)}=0$, or explicitly
\begin{eqnarray}
	\Delta+\alpha^{\prime\prime}\gamma=0,\qquad \gamma  =\int_{0}^{\pi} \cos(2x) |\varphi_{nk}(x) |^2dx.
	\label{cond_zero}
\end{eqnarray}
This  requirement for the eigenvalues to be real either can not be satisfied or can be satisfied only for a single gap edge [since  $\gamma$ depends on the indexes $(n,k)$ which are omitted for the sake of brevity of notations]. Thus in a general situation $\Delta\neq 0$. If $\Delta<-\alpha^{\prime\prime}\gamma$, then $E_{nk}^{(1)}$ describes effective dissipation, what means that the zero solution ($\psi\equiv0$) is an attractor. If however $\Delta>-\alpha^{\prime\prime}\gamma$, the zero solution becomes unstable and one should expect emergence of nonlinear coherent structures, which of course are possible if the respective linear pump is compensated by the nonlinear dissipation. This is precisely the case of $\sin\phi>0$ in the parametrization chosen above [see also Eq.(\ref{eq:N0})].

For $\Delta=0$ (this is the situation explored in all numerical simulations reported in this paper) the above arguments mean that the nonlinear stable structures are expectable only for $\gamma>0$. Then nonlinear patterns are generated form arbitrarily small initial fluctuations (see also Fig.~\ref{fig:gen-dis} below). Meantime the sign of $\gamma$ is not uniquely defined and depends on the band-gap edge. For the  cos-like lattice, explored here, positive $\gamma$ is verified only for the lowest bands. Hence, only close to the lowest band edges one can expect  to obtain low-density stable nonlinear patterns. This conjecture is verified for all numerical simulations performed in the present paper.

Finally, we observe the validity  of the  inequality
\begin{eqnarray}
\label{gamma}
|\gamma|<1
\end{eqnarray}
 which will be used for the discussion   in what follows.

\subsection{Multiple-scale analysis of the stability}

By varying the pump on can {\em selectively excite periodic patterns}, whenever more than one stable state  exist. In the low-density limit, i.e. at $N\to 0$, the stability of the periodic solutions can be tested using the multiple-scale analysis, similarly to the approach successfully used in the conservative case~\cite{KS}. Now $\varepsilon$ has to be used as a small parameter of the problem.
Following this approach, for the chosen band edge, say $E_n^{(\pm)}$, the wavefunction is approximated by $\psi\approx \varepsilon A (\xi,\tau)\varphi_{n}^{(\pm)} (x)\exp{(-iE_{n}^{(\pm)} t)}$ where $A(\xi,\tau)$ is an amplitude depending on slow variables $\xi=\varepsilon x$, $\tau=\varepsilon^2 t$, and $\varphi_{n}^{(\pm)}(x)$ is the Bloch wavefunction at the edge of the $n$-th band.
We observe, that by choosing $\varepsilon$ to be the small parameter  we implicitly impose the conditions where the characteristic scale of the excitations is determined by the complex part of the potential (contrary to case of a conservative system, where the small parameter of expansion is treated as a free parameter related to the detuning of the chemical potential towards the adjacent gap, see e.g.~\cite{KS}).

Using the standard algebra (the details can be found say in Ref.~\cite{KS}) one verifies that  $A(\xi,\tau)$ solves the dissipative nonlinear Schr\"odinger equation
\begin{eqnarray}
i  A_\tau =- (2M)^{-1} A_ {\xi \xi} +i\Lambda  A +e^{-i\phi}\chi  |A |^2A .
\label{eq:A}
\end{eqnarray}
Here $M  = (d^2E_n^{(\pm)}/dk^2)^{-1} $ is the effective mass,
\begin{eqnarray}
\label{lambda}
\Lambda =\alpha^\prime\gamma +\delta=\frac{\alpha^{\prime\prime}\gamma +\Delta}{\varepsilon^2}
\end{eqnarray}
is the effective gain (according to the analysis of the Sec.~\ref{sec:lin_spec} nonlinear patterns emerge near  edges with $\Lambda>0$, the conclusion which is also confirmed by the next analytical consideration and numerical simulations, what justifies the definition of $\Lambda$ as a gain),
\begin{equation}
\label{chi}
\chi  =\int_{0}^{\pi} |\varphi_n^{(\pm)} |^4dx
\end{equation}
 is the effective nonlinearity, and $\gamma$ is defined in (\ref{cond_zero}) with $\varphi_{nk}=\varphi_n^{(\pm)}$.
Within the framework of the  multiple-scale expansion  the condition (\ref{eq:N0}) for conservation of the number of atoms  can be recast as follows
\begin{equation}
\Lambda-\sin(\phi)\chi|A|^2=0.
\label{eq:NAconst}
\end{equation}

Eq.~(\ref{eq:A}) possesses a stationary, coordinate independent, solution in the form
\begin{eqnarray}
 A_{st} =\sqrt{\frac{\Lambda}{\chi \sin(\phi)}}e^{-i\Omega \tau}.
\label{eq:Ast}
\end{eqnarray}
 Here  $\Omega=\Lambda/\tan(\phi)$ is a constant determining the shift of the chemical potential outwards the gap edge: recall that now $\mu=E^{(\pm)}_n+\varepsilon^2\Omega$. Unlike in the case of conservative systems, the value of $\Omega$ is fixed. It is determined by the linear density of atoms and by the nonlinear dissipation controlled by the parameter $\phi$ through the relations
\begin{eqnarray}
N=\frac{ \alpha^{\prime\prime}\gamma+\Delta}{\chi\sin\phi},\qquad
\mu=E_n^{(\pm)}+\frac{\alpha^{\prime\prime}\gamma+\Delta}{\tan(\phi)}.
\label{eq:Neq}
\end{eqnarray}

Several important conclusions follow immediately from  Eqs.~(\ref{eq:Neq}). First, the limit $N\to 0$ can be reached only when the condition (\ref{cond_zero}) is satisfied, i.e. only subject to the requirement for the spectrum to be real. In this limit $\mu\to E_n^{(\pm)}$. For the general case one must have $N>0$ what leads us to the  condition on the linear dissipation $\alpha^{\prime\prime}\gamma+\Delta>0$ necessary for existence of the nonlinear periodic solutions. This constrain, which in Sec.~\ref{sec:lin_spec} was obtained form the analysis of the linear spectrum, is equivalent to the requirement $\Lambda>0$ in Eq.(\ref{eq:A}), and under the chosen parametrization of nonlinearity phase $\phi$ is necessary for the  particle conservation    (\ref{eq:NAconst}). In the original notations, the obtained necessary condition reads $\Gamma>(1-\gamma)\alpha''$ and means that the pump should be stronger than the dissipation $\alpha''$ reduced by the lattice factor $(1-\gamma)$ [notice that due to (\ref{gamma}) this is a positive factor]. Now the number of particles per one OL period cannot be smaller than the minimal value $N_{min}=(\alpha^{\prime\prime}\gamma+\Delta)/\chi$, which is  the number of particles in the case of pure dissipative nonlinearity, i.e. when ($\phi=\pi/2$). In this limit  $\mu\to E_n^{(\pm)}$.

As the second relevant property of the system as hand, following from (\ref{eq:Neq}) we emphasize that the chemical potential is not a free parameter, as this would happen in a conservative system, but is determined by the balance of incoming and dissipating atoms.  This balance can be controlled say by the nonlinear dissipation, parametrized by $\phi$. Respectively, obtaining solutions corresponding to different branches  (see e.g. Fig.~\ref{fig:N-E}, panels b and d) require change of the nonlinear dissipation. In practical terms this can achieved, say, by using variations of the light with a frequency close to a resonance with one of the  excited atomic state. Then the control parameter between the real and imaginary parts of the scattering length is the relation between the frequency detuning $\tilde{\delta} $ form the excited level and  the natural line width $\tilde{\Gamma}$: $\tilde{\delta}/ \tilde{\Gamma}$~\cite{nonlin-dissip_Fedichev}, and practically any value of the parameter $\phi$ is achievable (see, e.g., the examples considered in~\cite{nonlin-dissip_Fedichev}).

As it is customary, to check the stability of constant amplitude solutions, we consider small perturbations of the stationary solution:
$ A_{st} +\left( a e^{ik\xi-i\omega \tau}+\overline{b} e^{-ik\xi+i\overline{\omega} \tau}\right)e^{-i\Omega \tau}
$
(here $|a|,|b| \ll |A_{st} |$).
Substituting this ansatz in Eq.~(\ref{eq:A}) and linearizing with respect to $a$ and $b$, we obtain a dispersion relation  in the long-wavelength limit ($k \ll 1$)
\begin{eqnarray}
\omega=-i\Lambda  \pm\sqrt{ \Omega  k^2/M  -  \Lambda^2 }.
\label{eq:omega}
\end{eqnarray}
Thus for the stability of the background $A$ we have to require two conditions:
\begin{eqnarray}
\label{condition}
\Lambda  >0 \quad\mbox{ and}\quad  (\mu-E_n^{(\pm)})  M>0.
\end{eqnarray}
The first of these inequalities has simple meaning of the positive effective dissipation which must compensate the nonlinear losses. The second condition is of "conservative origin" and coincides with the stability of Bloch states in the conservative systems (see, e.g.,~\cite{KS}).

 Besides the stationary solution (\ref{eq:Ast}), Eq.(\ref{eq:A}) possesses a time-dependent homogeneous   solution
\begin{equation}
A(\tau)=a(\tau)\exp{\left(-i\cos(\phi)\chi\int_{\tau_0}^\tau{a^2(\tau^\prime)d\tau^\prime}\right)},
\label{eq:Anonst}
\end{equation}
where
\eqb
a(\tau)=
 \frac{\sqrt{\Lambda} a_0  }{\sqrt{\Lambda e^{2\Lambda(\tau_0-\tau )}+\chi\sin(\phi)a^2_0 \left(1-e^{2\Lambda(\tau_0-\tau )}\right)}}\nonumber
\eqe
$a_0=a(\tau_0)$, and $\tau_0$ is the initial time.

 For a positive $\Lambda$ and nonzero $a_0$ the distribution (\ref{eq:Anonst}) at $\tau \to \infty$ converges  to the stationary solution (\ref{eq:Ast}). In other words, the  pattern (\ref{eq:Ast}) is an attractor for  any nontrivial initial distributions, corresponding to a smoothly modulated Bloch function. The respective  time-dependence of the particle number per period can be represented as
\begin{equation}
N =
\frac{N_0(\alpha^{\prime\prime}\gamma+\Delta)}{ N_0\chi\sin\phi +e^{2(\alpha^{\prime\prime}\gamma+\Delta)(t_0-t)}\left(\alpha^{\prime\prime}\gamma+\Delta- N_0\chi\sin\phi \right)},
\label{eq:N-t-pos}
\end{equation}
where $N_0$ is the initial number of particles, and we restored to the original dimensionless variables.

Below we will also explore the dynamical regimes when after some interval of time the pump is switched off, i.e. $\Gamma=0$ and $\Delta=-\alpha^{\prime\prime}$ (respectively, $\delta=-\alpha^\prime$). Recalling the definition (\ref{lambda}) as well as the property (\ref{gamma}), we conclude that this is the case of $\Lambda=\alpha^{\prime}(\gamma-1)<0$. Now the particle conservation condition (\ref{eq:NAconst}) can not be met for the dissipative nonlinearity $\sin(\phi)>0$. It is evident, that in this case nonstationary solution (\ref{eq:Anonst}) evolves towards zero, i.e. the zero solution becomes an attractor. The
 respective dependence of the number of particles per period on time is now expressed as
\begin{equation}
N(t)=
\frac{N_0\alpha^{\prime\prime}(1-\gamma) e^{2\alpha^{\prime\prime}(\gamma-1)(t-t_0)}}{\alpha^{\prime\prime}(1-\gamma)+N_0 \chi\sin(\phi) \left(1-e^{2\alpha^{\prime\prime}(\gamma-1)(t-t_0)}\right)}.
\label{eq:N-t-neg}
\end{equation}
At $t-t_0\gg\left[\alpha^{\prime\prime}\right]^{-1}$ this decay has the exponential asymptotic
\eqb
N(t)\approx
\frac{N_0\alpha^{\prime\prime}(1-\gamma) }{\alpha^{\prime\prime}(1-\gamma)+\chi\sin(\phi)N_0}e^{2\alpha^{\prime\prime}(\gamma-1)(t-t_0)}.
\eqe

\section{Numerical study of coherent structures}
\label{sec:numerics}

\begin{figure}
  \begin{center}
   \begin{tabular}{c}
       \includegraphics{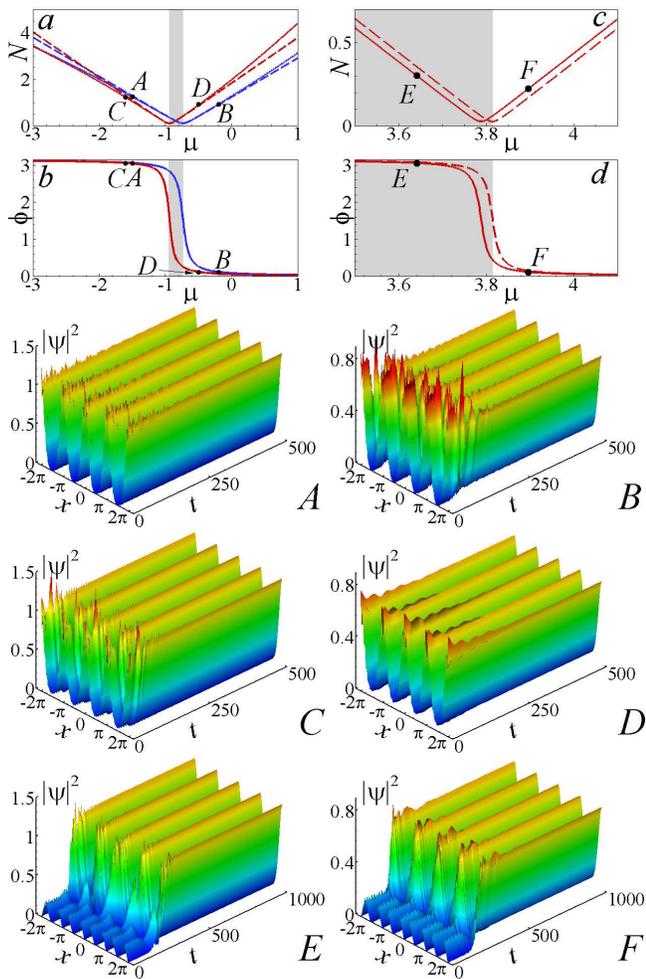}
   \end{tabular}
   \end{center}
\caption{(Color online) The number of particles $N$ (panels a,c) and the phase of the nonlinearity, $\phi$, (panels b,d) \textit{vs} the chemical potential $\mu$ for $\alpha^\prime=3.0$, $\alpha^{\prime\prime}=0.1$, $\Delta=0.0$ obtained analytically (dashed lines) from Eqs.~(\ref{eq:Neq}), and numerically (solid lines). Grey strips represent the first (in panels a,b) and the second  (in panels c,d) bands. The lower panels A--F illustrate temporal evolution of periodic solutions corresponding to the respectively labeled points of upper panels. The points A, C, and E are chosen to have the same nonlinearity phase $\phi\approx 3.055$. Similarly, the points B,D, and F are  characterized by the phase $\phi\approx 0.116$. The corresponding parameters $\gamma$ are $\gamma=0.641$ (panels A,B), $\gamma=0.536$ (panels C,D) , $\gamma=0.123$ (panels E,F). }
\label{fig:N-E}
\end{figure}

When the external parameters of the system (i.e. $\alpha^\prime$, $\alpha^{\prime\prime}$, $\Delta$ and $\phi$) are fixed, with each band-edge one can associate only one solution, having the number of particles and the chemical potential determined from (\ref{eq:Neq}) and varying when either of the parameters is changed. Bearing in mind physical applications of the model we concentrate on the study of the dependence of the number of particles on the chemical potential $N(\mu)$ [what is the same as $N(\Omega)$ due to the link between $\Omega$ and $\mu$]. The respective analysis was performed for the lowest bands and is summarized in Fig.~\ref{fig:N-E}.

The first lowest band ($n=1$) is associated to two branches of periodic solutions of Eq.~(\ref{eq:NSE}) depicted in Fig.\ref{fig:N-E} (panels a and b). One branch is $2\pi$-periodic  (the branch A--B) having zeros of the density in maxima of the OL in Eq.~(\ref{eq:NSE-0}), i.e. $x_p=\pi/2+p\pi$ (where $p$ is an integer). For this branch the dependence $N(\mu)$ reaches its minimum in the vicinity of  $E_1^{(+)}$. The second branch is $\pi$-periodic (the branch C--D). It has no zeros in the whole space, this reflecting the fact the respective linear Bloch wave of the underline conservative system has no zeros. The dependence $N(\mu)$ of the second branch reaches minimum at $E_1^{(-)}$. We notice, that analytical expressions for  $N(\mu)$ and $\phi(\mu)$ in the vicinity of $E_1^{(\pm)}$ remarkably well match the numerical ones (respectively dashed and solid lines in Fig.~\ref{fig:N-E}a,b).

Turning to the upper edges, in Fig.~\ref{fig:N-E}c,d we show the branch E--F, which is associated with the  upper edge of the second band. It has the minimum of the number of particles in the vicinity of $E_2^{(+)}$ given by $N_{min}\approx 0.026$, i.e. smaller than the one in the vicinity of first band edges ($N_{min}\approx 0.1$, see Fig.\ref{fig:N-E}a).  We also clearly observe that the minimum of $N(\mu)$ is considerably shifted to the lower values of $\mu$ in comparison with the analytical prediction (\ref{eq:Neq}).

Passing to the stability of the solutions, direct numerical integration of Eq.~(\ref{eq:NSE}) reveals that  $2\pi$-periodic patterns are stable at $\mu<E_1^{(+)}$ (solution A in Fig.\ref{fig:N-E}a,b) and unstable at $\mu>E_1^{(+)}$ (solution B in Fig.~\ref{fig:N-E}a,b). As this is typical for dissipative systems, the stable solution is an attractor, and therefore an unstable initial condition rapidly evolves towards the stable one having the same nonlinear phase $\phi$.  This process is illustrated in the panel B of Fig.\ref{fig:N-E}, where the unstable solution transforms into the attractor corresponding to the point D (c.f. panels A and B in Fig.~\ref{fig:N-E}). At the same time $\pi$-periodic solutions are stable at $\mu>E_1^{(-)}$ (see, e.g., solution D in Fig.\ref{fig:N-E}a,b and respective panel D) and unstable at $\mu<E_1^{(-)}$ (see e.g., solution C in Fig.~\ref{fig:N-E}a,b and respective panel C,  where we show the convergence of the unstable solution to the attractor corresponding to the point A).

The described stability properties  well corroborate with the simple analysis performed in the preceding sections.
This is not the case, however, for the solutions associated with the upper band edges. They appear to be dynamically unstable, again rapidly converging to the stable lower branches having the same parameter $\phi$. This is illustrated by the evolution of the patterns corresponding to the points E and F shown in the respective panels of Fig.~\ref{fig:N-E},  where they transform into the solutions, corresponding to points A and D, respectively. This however should not be viewed as the discrepancy with the above analysis of the stability. Indeed the latter one is done for slowly varying perturbations  thus representing only necessary but not yet sufficient condition for the stability. Moreover the reported behavior well agrees with the conclusion about emergence of complex eigenvalues due to weak dissipation [see (\ref{cond_zero}) as well as the subsequent discussion].
\begin{figure}
  \begin{center}
   \begin{tabular}{c}
       \includegraphics{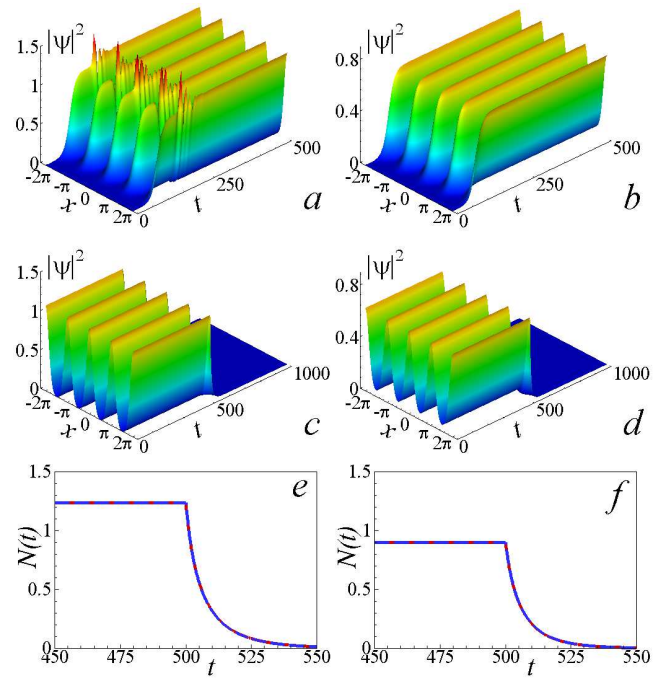}
   \end{tabular}
   \end{center}
\caption{ (Color online)   (a,b) The evolution of the density, obtained from the numerical integration of  Eq.~(\ref{eq:NSE}) with the initial condition $\psi(x,0)\equiv 0.01$. (c,d) The evolution of the density started with the stable patterns   corresponding to the points A and D of Fig.~\ref{fig:N-E} with the gain  turned off at   $t_0=500$. (e,f)  Evolution of the number of particles per one period corresponding to the dynamics shown in panels c,d, respectively,  calculated  numerically (solid curves) and analytically from Eq.~(\ref{eq:N-t-neg}) (dashed curves). Notice, that solid and dashed lines are indistinguishable in the scale of panels e,f. The other  parameters are $\alpha^\prime=3.0$, $\alpha^{\prime\prime}=0.1$, $\Delta=0.0$, $\phi\approx 3.055$ (panels a,c,e) and $\phi\approx 0.116$ (panels b,d,f). }
\label{fig:gen-dis}
\end{figure}

Having discussed the stability properties of the periodical patterns we address the two natural questions: how one can generate the above-mentioned stable periodical patterns, and how they evolve if after some time the gain is off? The answer to the first question is rather simple, since the final pattern is an attractor it will be generated from any initial distribution having small density. This is illustrated in Fig.~\ref{fig:gen-dis}a,b  where one observes evolution of a homogeneous nonzero excitation towards the \textit{stable} periodic pattern, whose parameters (number of particles per OL period and chemical potential) are determined by the fixed  phase $\phi$ of the nonlinearity. More specifically, in Fig.~\ref{fig:gen-dis}a the initial excitation is transformed into the solution, corresponding to point A in Fig.~\ref{fig:N-E}, while the initial excitation in Fig.~\ref{fig:gen-dis}b evolves towards the solution with the  parameters of the point D in Fig.~\ref{fig:N-E}. It is worth to notice, that such a spatio-temporal evolution confirms the behavior, qualitatively predicted by Eq.~(\ref{eq:Anonst}).

 The answer to the second of the above questions becomes evident from Fig.~\ref{fig:gen-dis}c-f, where we show the evolution of the stable periodic patterns, after the gain is turned off (recall that turning off the gain is equivalent to setting $\Delta=-\alpha^{\prime\prime}$). Thus, after turning off the gain, the condensate starts losing the atoms and its density homogeneously goes to zero [as it was predicted by Eq.~(\ref{eq:N-t-neg})].  The panels  e and f of Fig.~\ref{fig:gen-dis}  demonstrate remarkably  good correspondence between the numerical results and the analytical predictions.

\section{Coherent structures with quintic dissipation}
\label{sec:quintic}

Significant dissipative losses in a BEC can occur due to the inelastic three-body interactions~\cite{FMS,nonlin-dissip_Fedichev,nonlin-dissip}. Therefore as the next step we address their effect on emergence of the coherent structures in the dissipative model
\begin{eqnarray}
\label{eq:NSE-quint}
i\Psi_t=- \Psi_{xx}+2\alpha \sin^2(x)\Psi
+g^\prime |\Psi|^2\Psi+i\Gamma\Psi
\nonumber\\
  -ig^{\prime\prime}|\Psi|^4\Psi,
\end{eqnarray}
where $g^{\prime\prime}>0$, and other parameters are defined as above.
Similarly to the previous case, introducing the parameters $\Delta=\Gamma-\alpha^{\prime\prime}$, $g=g^{\prime}/{\sqrt{g^{\prime\prime}}}$,
and the renormalized macroscopic wave-function $\displaystyle{\psi =\left(g^{\prime\prime}\right)^{ 1/4}e^{-i\alpha^\prime t}\Psi }$, we arrive at the equation
\begin{eqnarray}
i\psi_t=- \psi_{xx}-(\alpha^{\prime}-i\alpha^{\prime\prime})\cos(2x)\psi+i\Delta\psi
\nonumber\\
+g|\psi|^2\psi-i|\psi|^4\psi\,.
\label{eq:NSE-quint-new}
\end{eqnarray}
Now  the conservation of the  number of atoms requires [c.f. Eq.~(\ref{eq:N0})]
\begin{equation}
\int_{0}^{\pi} \left[\alpha^{\prime\prime}\cos(2x)  |\psi |^2-|\psi |^6\right] dx+\Delta N=0.
\label{eq:N0-quint}
\end{equation}

Following the approach developed above for the analysis of the   cubic dissipative term, now we  employ the multiple-scale analysis of Eq.~(\ref{eq:NSE-quint-new}) representing the wavefunction in the form $\psi\approx \varepsilon^{1/2} A(\xi,\tau)\varphi^{\pm}(x)\exp{(-iE^{(\pm)}_nt)}$. The slowly varying amplitude $A$ now is governed by the equation
\begin{eqnarray}
i  A_\tau =- (2M)^{-1} A_ {\xi \xi} +i\Lambda  A +G\chi  |A |^2A-i\Upsilon  |A |^4A , \label{eq:A-quin}
\end{eqnarray}
where  $\Upsilon =\int_{0}^{\pi} |\varphi^{\pm} |^6dx$, $g=\varepsilon G$ ($\Upsilon\sim|G|\sim 1$) and other parameters are defined as in (\ref{eq:A}). We emphasize that unlike in the cubic case, here we imposed the condition of smallness of the conservative nonlinearity coefficient $g$, since in this case the effects of the elastic two-body interactions and the inelastic three-body interactions become of the same order and must be accounted simultaneously.

\begin{figure}
  \begin{center}
   \begin{tabular}{c}
       \includegraphics{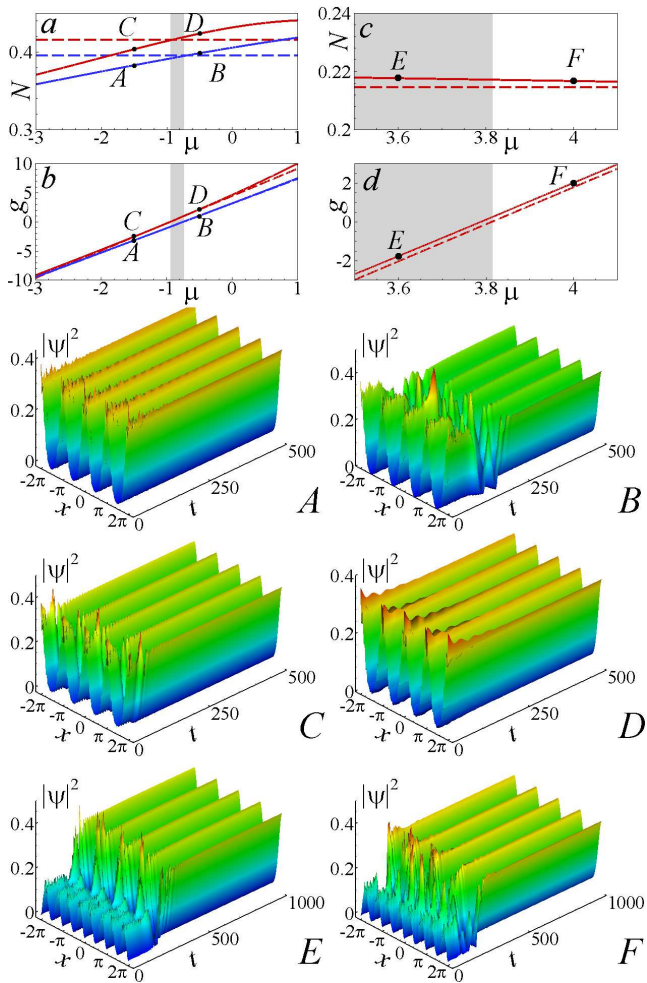}
   \end{tabular}
   \end{center}
\caption{(Color online) Number of particles $N$ (panels a,c) and cubic nonlinearity $g$ (panels b,d) \textit{vs} chemical potential $\mu$ for $\alpha^\prime=3.0$, $\alpha^{\prime}=0.1$, $\Delta=0.0$ calculated analytically (dashed lines) from Eqs.(\ref{eq:Neq-quintic}), or numerically (solid lines). Grey strips represent the first (in panels a,b) and the second (in panels c,d) bands. In the lower panels A--F we show  temporal evolution of periodic solutions at respective points of upper panels. }
\label{fig:quin}
\end{figure}

Now the constant-amplitude solution reads
\begin{equation}
A_{st}=(\Lambda/\Upsilon)^{1/4}e^{-i\Omega \tau}.
\label{eq:Ast-quin}
\end{equation}
The number of particles per OL period $N$ and the chemical potential $\mu$ depend on the nonlinearity $g$
\begin{eqnarray}
N=\sqrt{\frac{\alpha^{\prime\prime}\gamma+\Delta}{\Upsilon}},\qquad
\mu=E_n^{(\pm)}+g\chi  \sqrt{\frac{\alpha^{\prime\prime}\gamma+\Delta}{\Upsilon}}.
\label{eq:Neq-quintic}
\end{eqnarray}
As it can be seen from Eqs.~(\ref{eq:Neq-quintic}), in the case of dissipative quintic nonlinearity the analytically estimated density of particles in a stationary mode does not depend upon the cubic nonlinearity $g$ [unlike this happens in the cubic case, see (\ref{eq:Neq})]. The two-body interactions, however, determine
the chemical potential $\mu$ for given $N$. In particular, since $\chi>0$ [see the definition (\ref{chi})] we have that  $\mu<E_n^{(\pm)}$ ($\mu>E_n^{(\pm)}$)
when $g<0$ ($g>0$). Fig.~\ref{fig:quin} illustrates that the simple analytical estimate for $N(\mu)$ (shown in panels a,c by dashed lines) is in agreement with the numerical values only in the vicinity of the band edges (unlike this was in the case of the cubic dissipative nonlinearity where the domain of quantitative coincidence of the analytical and numerical results was significantly larger, see Fig.~\ref{fig:N-E}). The analytical estimates for  $g(\mu)$ is in much better agreement with the dependence found numerically  (Fig.~\ref{fig:quin}b,d).

Turning to the simple stability analysis of the dissipative Bloch waves, first we notice that the analysis performed in Sec.~\ref{sec:lin_spec} continues to be valid. Moreover, analyzing the stationary solutions (\ref{eq:Ast-quin}) we arrive at the conditions (\ref{condition}),  earlier   obtained for the cubic case. Also similarly to  the cubic case,  direct numerical simulations confirm  the validity of (\ref{condition}) for the first band (see Fig.~\ref{fig:quin}a,b, where modes A and D are stable while modes B and C are unstable) and their failure for the top of the second band (see Fig.~\ref{fig:quin}c,d showing that modes E and F are unstable).

\section{Associated dissipative lattices}
\label{sec:lattice}

As the last point, we observe, that the limit of a relatively large potential depth can be  described by the tight-binding approximation which is obtained by discretizing Eq.~(\ref{eq:NSE})  using the expansion
\begin{eqnarray}
\psi(x,t)= \sum_{nk} c_{nk}(t)w_{nk}(x)
\label{Wannier}
\end{eqnarray}
 over the basis of the
Wannier functions
\begin{eqnarray}
w_{nm}(x)=\frac{1}{\sqrt{2}}\int_{-\infty}^\infty\varphi_{nq}(x)e^{-i\pi m q}dq.
\end{eqnarray}
For more precise conditions of the validity of the approximation as well as for the details of the expansion we refer to~\cite{AKKS}, only mentioning here that it must be required sufficiently fast convergence of the coefficients $\omega_{nm}$ of the Fourier expansion of the energy:
\begin{eqnarray}
\omega_{n0} \gg \omega_{n1}
\gg\cdots,
\quad \omega_{nm}=\frac 12
\int_{-1}^{1} E_n(q)e^{-i\pi mq }dq
\end{eqnarray}
allowing one to restrict the consideration to the hopping of only nearest neighbors.

At this point however we emphasize one important feature which distinguishes the Wannier function mapping of the continuous model to a discrete one in the dissipative case. While, in general, the tight-binding limit does not account a number if important features of the dynamics of the underline continuum model~\cite{AKKS}, the tight binding approximation can be formally performed for a number of the lowest bands (provided the depth of the potential is high enough), still giving reasonable approximation for the static solutions. In the dissipative case  the situation is dramatically changed because of the stability properties. Namely, now the tight-binding approximation makes sense only for a band where the periodic solutions are attractors. Even when other bands result in discrete dissipative models with stable solutions, difference  between such solutions and those of the original continuum model   grows exponentially already at early stages of the evolution. Therefore, in what follows we assume that the  expansion (\ref{Wannier}) is performed for the band associated to the stable solutions. Moreover, in accordance with the examples considered above we choose the first lowest band with $n=1$. This assumption allows us  to drop the band index $n$ below.

Then, using the orthogonality of the Wannier functions we readily obtain the dissipative discrete NLS equation for the expansion coefficients $c_n$
\begin{eqnarray}
i\frac{dc_m}{dt}=\left(1-i\frac{\alpha^{\prime\prime}}{\alpha^{\prime}}\right)\left(\omega_0 c_m+\omega_1c_{m+1}+\omega_1c_{m-1}\right) \nonumber\\
+e^{-i\phi}W|c_m|^2c_m.
\label{eq:disc}
\end{eqnarray}
Here $W=\int_{-\infty}^{\infty} w_{1,m}^4(x)dx$ and we have neglected hopping integrals with the upper bands and among next nearest neighbors. The obtained equation (\ref{eq:disc}) was considered earlier in \cite{c:ErfChris2003} (it is also relevant to mention studies of a more general discrete dissipative model in~\cite{Abdul}).
Therefore we do not proceed with the further analysis of  (\ref{eq:disc}), referring for the study of their modulational instability to the mentioned work.

 We complete this last section with the indication that the qubic-quintic dissipative model (\ref{eq:NSE-quint-new}) can be discretized in the similar manner
\begin{eqnarray}
i\frac{dc_m}{dt}=\left(1-i\frac{\alpha^{\prime\prime}}{\alpha^{\prime}}\right)\left(\omega_0c_m+\omega_1c_{m+1}+\omega_1c_{m-1}\right)\nonumber\\
+gW|c_m|^2c_m-iB|c_m|^4c_m,
\label{eq:disc_1}
\end{eqnarray}
where $B=\int_{-\infty}^{\infty} w_{1,m}^6(x)dx$. This model was also addressed in \cite{c:ErfChris2003}.

\section{Conclusions}

To conclude, we have reported emergence of nonlinear periodic structures in a nonlinear Schr\"odinger equation with a dissipative periodic potential, nonlinear losses and linear pump. The models describe a Bose-Einstein condensate loaded in an optical lattice, where the potential created by the light interaction with the condensate accounts for nonelastic interactions of photons with atoms. We addressed nonlinear losses due to two- and three-body interactions.    These simple one-dimensional model reveals existence of stable nonlinear Bloch waves (attractors) to which large range of the initial data converges. The obtained solutions do not have the linear limit (i.e. the limit of the zero density) and the number of particles is limited from below.
In all the simulations performed the observed stable periodic patterns correspond to the lowest band. The existence of the attractor has important physical consequence: in the described models loading particles (using the linear pump) is only possible to the given state, which is imposed by the nonlinear losses. By changing the interatomic interactions one can follow different branches of the stable solutions.

Meantime the simplest stability analysis performed in the present paper does not yet describe all the features of the system which  are related, in particular, to the scales comparable with the lattice constant, and thus not accounted by the multiple-scale expansion. These issues of the theory are left for further studies.

\acknowledgments

Authors are grateful to F. Kh. Abdullaev for useful comments.
Y.V.B. acknowledges partial support from FCT, Grant No. SFRH/PD/20292/2004. The research of VVK was partially supported by a Marie Curie International Incoming
Fellowship within the 7th European Community Framework Programme under the grant PIIF-GA-2009-236099 (NOMATOS).

\end{document}